\documentclass[aps,prd,preprint,nofootinbib]{revtex4}
\usepackage{amsfonts}
\usepackage{mathrsfs}
\usepackage{graphicx}
\usepackage{amsmath}
\usepackage{amssymb}
\usepackage{multirow}
\usepackage{subfigure}
\usepackage{epsfig}
\usepackage{graphicx}
\usepackage{booktabs}
\usepackage{array}
\usepackage{tabularx}

\newcommand{\bqa}{\begin{eqnarray}}
\newcommand{\eqa}{\end{eqnarray}}
\newcommand{\beq}{\begin{equation}}
\newcommand{\eeq}{\end{equation}}
\graphicspath{{fig/}{dia/}} \DeclareGraphicsExtensions{.eps}
\begin{document}
\baselineskip 20pt
\title{NLO QCD corrections to $J/\psi$ pair production in photon-photon collision}

\author{\vspace{1cm} Hao Yang$^1$\footnote[2]{
yanghao174@mails.ucas.ac.cn}, Zi-Qiang Chen$^{1}$\footnote[3]{
chenziqiang13@mails.ucas.ac.cn}  and Cong-Feng
Qiao$^{1,2}$\footnote[1]{qiaocf@ucas.ac.cn, corresponding author} \\}

\affiliation{$^1$ School of Physics, University of Chinese Academy of
Sciences, Yuquan Road 19A, Beijing 100049\\
$^2$ CAS Key Laboratory of Vacuum Physics, Beijing 100049, China\vspace{0.6cm}}

\begin{abstract}

We calculate the next-to-leading order (NLO) quantum chromodynamics (QCD) correction to the exclusive process $\gamma+\gamma\to J/\psi+J/\psi$ in the framework of non-relativistic QCD (NRQCD) factorization formalism.
The cross sections at the SuperKEKB electron-positron collider, as well as at the future colliders, like the Circular Electron Positron Collider (CEPC) and the International Linear Collider (ILC), are evaluated.
Numerical result indicates that the process will be hopefully to be seen by the Belle II detector within the next decade.

\vspace {5mm} \noindent {PACS number(s): 12.38.Bx, 12.39.Jh, 14.40.Pq, 14.70.Bh}
\end{abstract}

\maketitle

\section{INTRODUCTION}

The $J/\psi$ meson has been proposed as an ideal laboratory to investigate both the perturbative and non-perturbative properties of quantum chromodynamics (QCD) since its discovery in 1974.
The non-relativistic QCD (NRQCD) factorization formalism \cite{Bodwin:1994jh}, which proposed by Bodwin, Braaten, and Lepage, provide a systematic framework for the theoretical study of quarkonium physics.
In the NRQCD factorization formalism, the quarkonium production and decay rates can be factorized as the process dependent, but perturbative calculable short distance coefficients multiplied by the supposed universal long distance matrix elements (LDMEs).
The relative importance between the LDMEs can be estimated by means of velocity scaling rules.
In this way, the theoretical prediction takes the form of a double expansion in the strong coupling constant $\alpha_s$ and the heavy quark velocity $v$.

Quarkonium production in photon-photon collision is an interesting topic to study, where the signals are relatively clean.
At present, the inclusive $J/\psi$ production via photon-photon scattering was studied extensively \cite{Klasen:2001cu,Abdallah:2003du,Qiao:2003ba,Qiao:2003ba,Li:2009zzu,Butenschoen:2011yh}.
These researches indicate that the color-singlet (CS) sectors are inadequate to explain the experimental data \cite{Klasen:2001cu,Chen:2016hju}.
While the color-octet (CO) contributions based on different LDME sets diverse quite a lot, which even lead to opposite conclusions in some cases \cite{Klasen:2001cu,Butenschoen:2011yh}.
Besides the inclusive single $J/\psi$ production, the exclusive $J/\psi$ pair production, where the CO contributions are suppressed by $v^8$, also provides a test to the NRQCD factorization formalism at photon-photon colliders.
In Ref.\cite{Qiao:2001wv}, the leading order (LO) calculation of exclusive $J/\psi$ pair production via photon-photon fusion was performed within the NRQCD factorization framework, while the $J/\psi$ pair diffractive production was investigated in the pomeron exchange scheme \cite{Kwiecinski:1998sa,Baranov:2012vu,Carvalho:2015mra,Goncalves:2015sfy}.
Considering the fact that the next-to-leading order (NLO) QCD corrections to physical processes in quarknoium energy regime are normally significant \cite{Zhang:2005cha,Gong:2007db,Chen:2014lqa,Sun:2014gca}, in this work we calculate the $\gamma+\gamma\to J/\psi+J/\psi$ process at the NLO accuracy.

The rest of the paper is organized as follows.
In section II, we present the primary formulae employed in the calculation.
In section III, we elucidate some technical details for the analytical calculation.
In section IV, the corresponding numerical evaluation is performed.
The last section is reserved for summary and conclusions.

\section{FORMULATION}
In this work, we investigate the photon-photon scattering in $e^+e^-$ colliders.
The initial photons are assumed to be generated by the bremsstrahlung or the laser back scattering (LBS) effects.
The energy spectrum of bremsstrahlung photon can be well formulated in Weizsacker-Williams approximation (WWA) \cite{Frixione:1993yw} as
\begin{equation}
    f_{\gamma}(x) = \frac{\alpha}{2\pi}\left(\frac{1+(1-x)^2}{x}\log\left(\frac{Q^{2}_{\rm max}}{Q^{2}_{min}}\right)+2m^2_{e}x\left(\frac{1}{Q^{2}_{\rm max}}-\frac{1}{Q^{2}_{\rm min}}\right)\right),
\end{equation}
where $Q^{2}_{\rm min}=m^{2}_{e}x^{2}/(1-x)$ and $Q^{2}_{\rm max}=(\theta_{c}\sqrt{s}/2)^2(1-x)+Q^{2}_{\rm min}$, $x=\frac{E_{\gamma}}{E_{e}}$ is the energy fraction of photon, $\sqrt{s}$ is the collision energy for $e^+e^-$ collider, $\theta_{c}$ is the maximum scattering angle of the electron or positron.

The energy spectrum of LBS photon is \cite{Ginzburg:1981vm}
\begin{equation}
    f_{\gamma}(x)=\frac{1}{N}\left(1-x+\frac{1}{1-x}-4r(1-r)\right),
\end{equation}
where $r=\frac{x}{x_{m}(1-x)}$ and N is the normalization factor:
\begin{equation}
    N=\left(1-\frac{4}{x_{m}}-\frac{8}{x^{2}_{m}}\right)\log(1+x_{m})+\frac{1}{2}+\frac{8}{x_{m}}-\frac{1}{2(1+x_{m})^2}.
\end{equation}
Here $x_{m} \approx 4.83$ \cite{Telnov:1989sd} and the maximum energy fraction of LBS photon is restricted by $0 \leq x \leq \frac{x_m}{1+x_m}\approx 0.83$.

The total cross section can be expressed as the convolution between the cross section for $\gamma + \gamma \rightarrow J/\psi+J/\psi$ process and the photon distribution functions,
\begin{equation}
    d\sigma=\int dx_{1}dx_{2} f_{\gamma}(x_{1})f_{\gamma}(x_{2})d \hat{\sigma}( \gamma + \gamma \rightarrow J/\psi+J/\psi)\ ,
\end{equation}
where the $d \hat{\sigma}$ is calculated perturbatively up to the NLO level,
\begin{equation}
    d\hat{\sigma}( \gamma + \gamma \rightarrow J/\psi+J/\psi)=d\hat{\sigma}_{\rm born}+d\hat{\sigma}_{\rm NLO}+\mathcal{O}(\alpha^{2}\alpha^{4}_{s})\ .
\end{equation}
The born level cross section and the NLO correction take the following forms:
\begin{equation}
    \begin{split}
        &d\hat{\sigma}_{\rm born}=\frac{1}{2\hat{s}}\overline{\sum}|\mathcal{M}_{\rm tree}|^{2}d{\rm PS}_{2}\ ,\\
        &d\hat{\sigma}_{\rm NLO}=\frac{1}{2\hat{s}}\overline{\sum}2{\rm Re}(\mathcal{M}^{*}_{\rm tree}\mathcal{M}_{\rm oneloop})d{\rm PS}_{2}\ .
    \end{split}
\end{equation}
Here $\hat{s}$ is the center-of-mass energy square for the two photons, $\overline{\sum}$ means sum (average) over the polarizations and colors of final (initial) state particles, $d{\rm PS}_{2}$ denotes two-body phase space.

The standard form of quarkonium spin and color projection operator is adopted in our calculation \cite{Bodwin:2013zu}:
\begin{equation}
       v(\bar{p})\bar{u}(p)=\frac{1}{4\sqrt{2}E(E+m_c)}(\not\!\bar{p}-m_{c})\not\! \epsilon^{*}_{S}(\not\!P+2E)(\not\!p+m_{c})\otimes(\frac{\bf{1}_{c}}{\sqrt{N_{c}}})\ ,
\end{equation}
Here, $p$ and $\bar{p}$ are the momenta of heavy quark and antiquark respectively, $P=p+\bar{p}$ is the momentum of quarkonium, $E=\sqrt{P^{2}/4}$ is the energy of heavy (anti)quark in quarkonium rest frame, $\epsilon^{*}_{S}$ denotes the polarization vector, $\bf{1}_{c}$ represents the unit color matrix, and $N_c=3$ is the number of colors in QCD.
At the leading order of the relative velocity expansion, it is legitimate to take $p=\bar{p}=P/2$ and $E=m_c$.

\section{ANALYTICAL CALCULATION}

\begin{figure}[thbp]
    \begin{center}
    \includegraphics[scale=0.7]{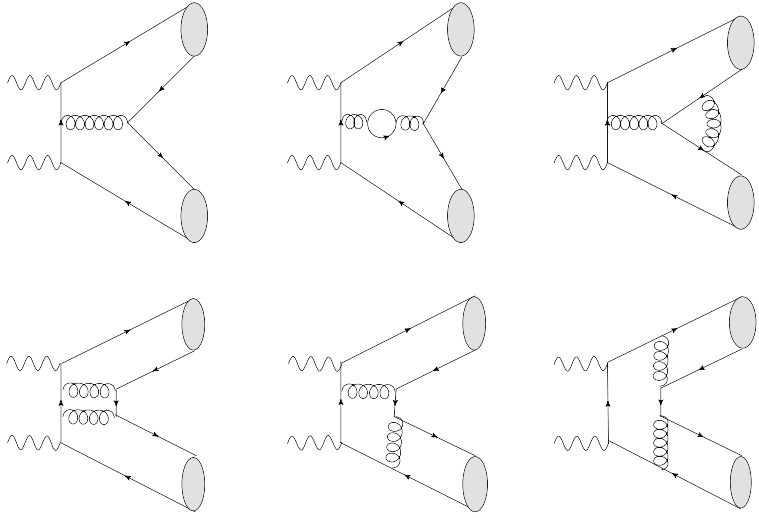}
    \caption{Typical Feynman diagrams for $\gamma+\gamma\rightarrow J/\psi+J/\psi$ process at LO and NLO . \label{fig1}}
    \end{center}
\end{figure}

The typical Feynman diagrams for $\gamma+\gamma\rightarrow J/\psi+J/\psi$ process are shown in Fig.\ref{fig1}.
The LO calculation is straightforward and the result is pretty simple.
We reproduce the Eq.(4) of Ref.\cite{Qiao:2001wv}, where the analytical expression for LO differential cross section was presented.

In the calculation of NLO correction, the ultraviolet (UV) and infrared (IR) singularities are regularized by the dimensional regularization with $D=4-2\epsilon$.
The IR singularities are canceled each other and the UV singularities are removed by renormalization procedure.
The relevant renormalization constants include $Z_{2}, Z_{3}, Z_{m}$ and $Z_{g}$, which corresponding to heavy quark field, gluon field, heavy quark mass and strong coupling constant, respectively.
Among them, $Z_{2}$ and $Z_{m}$ are defined in the on-mass-shell (OS) scheme, while others are defined in the  modified minimal-subtraction ($\overline{\rm MS}$) scheme. The counter terms are
\begin{equation}
    \begin{split}
        &\delta Z^{\rm OS}_{2}=-C_{F}\frac{\alpha_{s}}{4\pi}\left[\frac{1}{\epsilon_{\rm UV}}+\frac{2}{\epsilon_{\rm IR}}-3\gamma_{E}+3\ln\frac{4\pi\mu^{2}}{m_c^{2}}+4\right],\\
        &\delta Z^{\rm OS}_{m}=-3C_{F}\frac{\alpha_{s}}{4\pi}\left[\frac{1}{\epsilon_{\rm UV}}-\gamma_{E}+\ln\frac{4\pi\mu^{2}}{m_c^{2}}+\frac{4}{3}\right],\\
        &\delta Z^{\overline{\rm MS}}_{3}=(\beta_{0}-2C_{A})\frac{\alpha_{s}}{4\pi}\left[\frac{1}{\epsilon_{\rm UV}}-\gamma_{E}+\ln(4\pi)\right],\\
        &\delta Z^{\overline{\rm MS}}_{g}=-\frac{\beta_{0}}{2}\frac{\alpha_{s}}{4\pi}\left[\frac{1}{\epsilon_{\rm UV}}-\gamma_{E}+\ln(4\pi)\right].
        \label{eq_ctterm}
    \end{split}
\end{equation}
Here $\mu$ is the renormalization scale, $\gamma_{E}$ is the Euler's constant, $\beta_{0}=\frac{11}{3}C_{A}-\frac{4}{3}T_{F}n_{f}$ is the one-loop coefficient of the QCD $\beta$ -function, and $n_{f}$ is the active quark flavor numbers. The color factors $C_{A}=3, C_{F}=\frac{4}{3}$ and $T_{F}=\frac{1}{2}$.

In the calculation, the Mathematica package FeynArts \cite{Hahn:2000kx} is used to generate Feynman diagrams; FeynCalc \cite{Mertig:1990an} and FeynCalcFormLink \cite{Feng:2012tk} are used to handle the algebraic calculation;
The package FIRE \cite{Smirnov:2008iw} is employed to reduce all the one-loop integrals into typical master intergrals $A_{0}, B_{0}, C_{0}, D_{0}$, which can be numerically evaluated by the LoopTools \cite{Hahn:1998yk}.
The overall phase space integrals are performed numerically by using the package CUBA \cite{Hahn:2004fe}.

\section{NUMERICAL RESULTS}

In the numerical calculation, the general parameters are taken as $\alpha=\frac{1}{137.065}$, $m_{e}=0.511$ MeV and $m_{c}=1.5$ GeV.
The $J/\psi$ radial wave function at the origin is extracted from $J/\psi$'s leptonic width:
\begin{equation}
\Gamma(J/\psi\to e^+e^-)=\frac{4\alpha^2}{9m_c^2}|R_S(0)|^2\left(1-4C_F\frac{\alpha_s(\mu_0)}{\pi}\right).
\end{equation}
By taking $\mu_0=2m_c$ and $\Gamma(J/\psi\to e^+e^-)=5.55$ keV \cite{Tanabashi:2018oca}, we obtain $|R_S^{\rm LO}(0)|^2=0.528$ GeV$^3$ and $|R_S^{\rm NLO}(0)|^2=0.907$ GeV$^3$, which will be used in the LO and NLO calculation respectively.
The two-loop formula
\begin{equation}
    \frac{\alpha_{s}(\mu)}{4\pi}=\frac{1}{\beta_{0}L}-\frac{\beta_{1}\ln L}{\beta^{3}_{0}L^{2}},
\end{equation}
for the running coupling constant is employed in the NLO calculation, in which,  $L=\ln (\mu^{2}/\Lambda^{2}_{\rm QCD})$, $\beta_0=\tfrac{11}{3}C_A-\tfrac{4}{3}T_Fn_f$, $\beta_{1}=\frac{34}{3}C^{2}_{A}-4C_{F}T_{F}n_{f}-\frac{20}{3}C_{A}T_{F}n_{f}$, with $n_f=4$ and $\Lambda_{\rm QCD}=297$ MeV.
For the LO calculation, the one-loop formula for the running coupling constant is used.

In the near future, it turns out the $\gamma+\gamma\to J/\psi+J/\psi$ process is hopeful to be observed by Belle II detector at the SuperKEKB electron-positron collider due to its high luminosity.
The collider beam energies of the positron and electron are 4 GeV and 7 GeV respectively, which yields
a center-of-mass energy of 10.6 GeV. The corresponding LO and NLO cross sections for $\gamma+\gamma\to J/\psi+J/\psi$ process with the WWA photon as the initial state are shown in Table \ref{tab0}. 
The angular cut $\theta_c$ of the WWA is set to be 83 mrad, the crossing angle between the positron and electron beams \cite{Abudinen:2019osb}.
The transverse momentum cut $p_{t}\ge 0.01$ GeV is imposed on each individual $J/\psi$. 
In order to estimate the theoretical uncertainty, four reasonable choice for the renormalization scale \cite{Baranov:1997wy}, i.e. $\mu=2m_c$, $\mu=\sqrt{4m_c^2+p_t^2}$, $\mu=\sqrt{\hat{s}}/2$ and $\mu=\sqrt{\hat{s}}$ are taken, where $\sqrt{\hat{s}}$ is the center-of-mass energy of the photon-photon system.
It can be seen that the theoretical uncertainty induced by renormalization scale is even larger at NLO than that at the LO,
which seems a little counterintuitive and needs a further investigation.

\begin{table}[ht]
    \caption{The LO and NLO total cross sections for $J/\psi$ pair production via photon-photon fusion at the SuperKEKB  collider. Here, the transverse momentum cut $p_{t}\ge 0.01$ GeV is imposed on every single $J/\psi$.}
    \begin{center}
          \begin{tabular}{p{1.8cm}<{\centering} p{1.8cm}<{\centering} p{1.8cm}<{\centering} p{1.8cm}<{\centering} p{1.8cm}<{\centering}}
        \toprule
        \hline
            $\mu$  & $2m_c$  &  $\sqrt{4m_c^2+p_t^2}$ & $\sqrt{\hat{s}}/2$ & $\sqrt{\hat{s}}$ \\
        \hline
        $\sigma_{\rm LO}$(fb)&
        $0.558$ &
        $0.527$ &
        $0.503$&
        $0.305$  \\

       $\sigma_{\rm NLO}$(fb)&
        $0.100$ &
        $0.154$ &
        $0.194$&
        $0.399$  \\
        \botrule
      \end{tabular}
    \end{center}
    \label{tab0}
\end{table}

The instantaneous luminosity of the SuperKEKB collider may reach $8\times10^{35}\ {\rm cm}^{-2}{\rm s}^{-1}\approx 2.52\times 10^4\ {\rm fb}^{-1}{\rm year}^{-1}$ in 2025 \cite{urlsuperkekb}.
Hence we will obtain $(2.52\sim 10.05)\times 10^3$ $J/\psi$ pair events per year.
In experiment, $J/\psi$ can be reconstructed  through its leptonic decays.
According to Ref. \cite{Tanabashi:2018oca}, the branching ratio ${\rm Br}(J/\psi\to l^+l^-(l=e,\mu))=12\%$, then the number of candidates per year is $36\sim 145$. Considering the signals are relatively clean, the theoretical estimation is hopefully testable in Belle II experiment.

\begin{figure}[!thbp]
    \centering
    \subfigure[]{\includegraphics[width=0.49\textwidth]{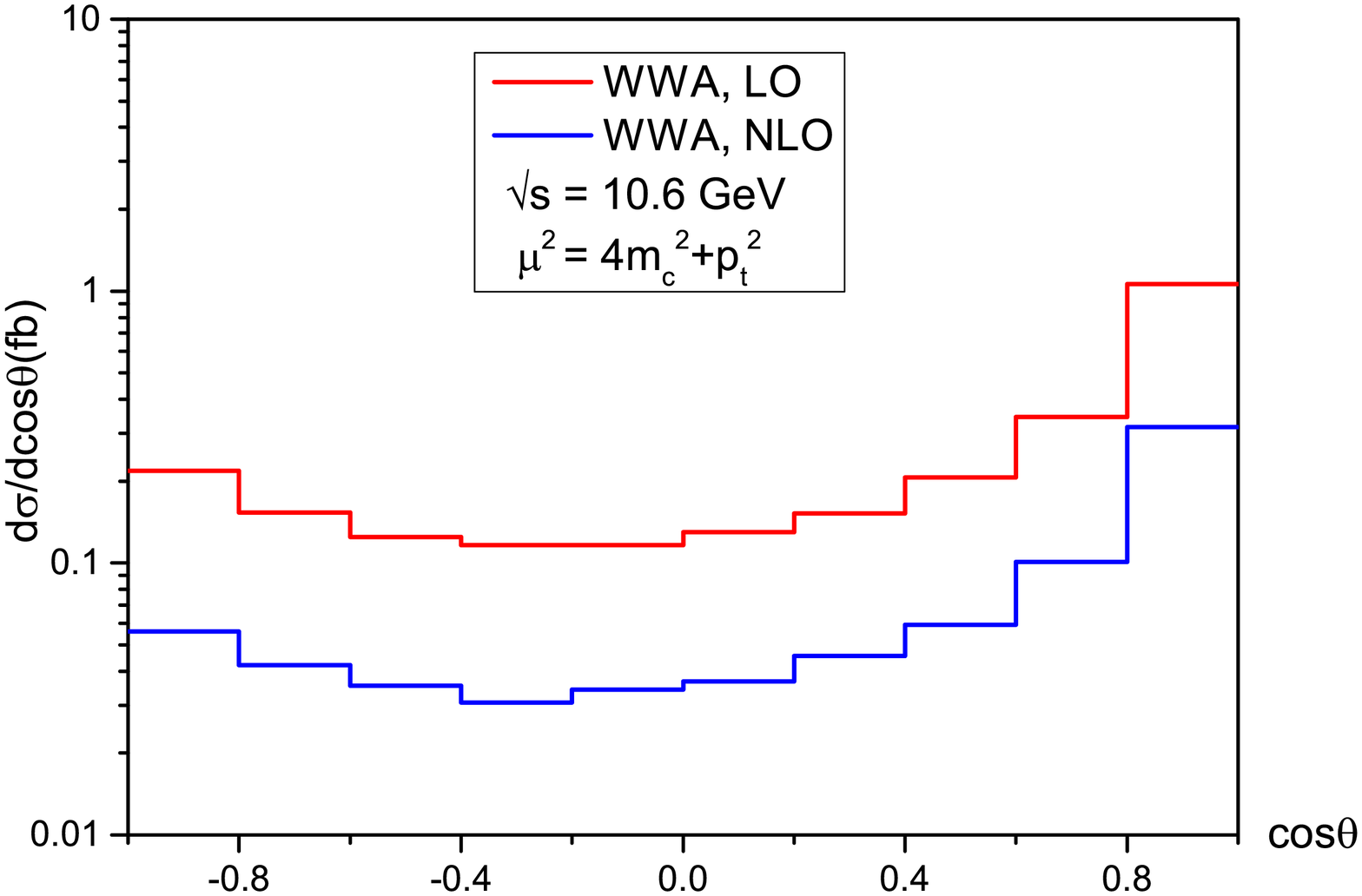}}
    \subfigure[]{\includegraphics[width=0.49\textwidth]{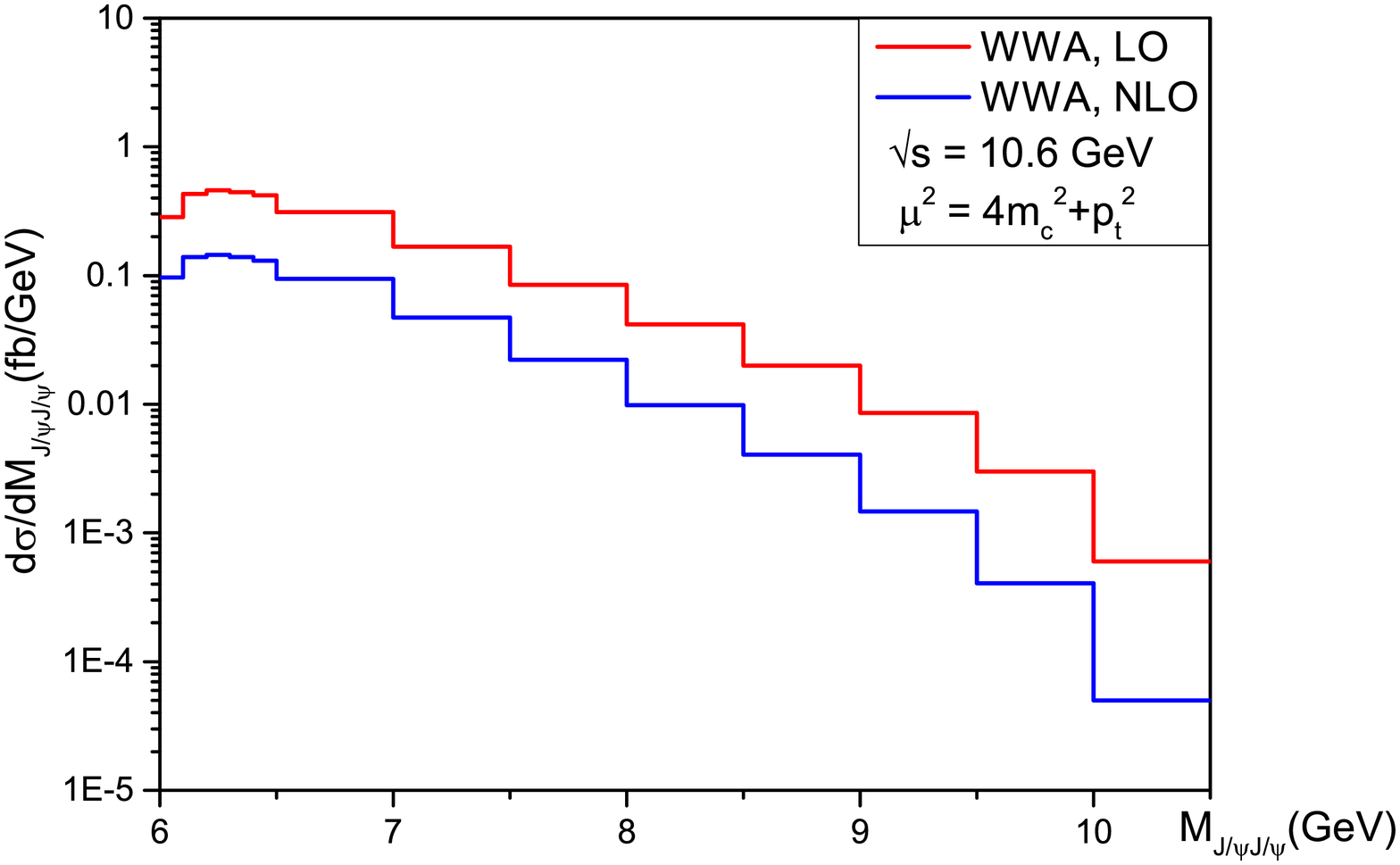}}
    \caption{The differential cross sections in bins of: (a) ${\rm cos}\theta$, (b) $M_{J/\psi J/\psi}$. Here the renormalization scale $\mu=\sqrt{4m_{c}^2+p^{2}_{t}}$. The transverse momentum cut $p_{t}\ge 0.01$ GeV is imposed on each single $J/\psi$.\label{fig2}}
\end{figure}

The differential cross sections $d\sigma/d{\rm cos}\theta$ at the SuperKEKB collider are shown in Fig. \ref{fig2}(a), where $\theta$ is the scattering angle. It can be seen that the events are more likely to be produced along the electron beam direction, which is one of the consequences of the asymmetry beams energy of the SuperKEKB collider.
The differential cross sections $d\sigma/dM_{J/\psi J/\psi}$ are shown in Fig. \ref{fig2}(b), where $M_{J/\psi J/\psi}=\sqrt{\hat{s}}$ is invariant mass of the $J/\psi$ pair. The invariant mass distribution arrives at the peak near threshold energy.

In the future $e^+e^-$ colliders, like the Circular Electron Positron Collider (CEPC) \cite{CEPCStudyGroup:2018ghi} and the International Linear Collider (ILC) \cite{Baer:2013cma}, the collision energy may reach 250 GeV or 500 GeV.
And the LBS photon collision can be realized by imposing a laser beam to each $e$ beam.
Therefore, we investigate the $J/\psi$ pair production under both WWA and LBS photon collision with $\sqrt{s}=250$ GeV and $\sqrt{s}=500$ GeV.
The corresponding LO and NLO total cross sections are show in Table \ref{tab1}.
The angular cut $\theta_c$ in WWA is set to be 32 mrad, as at the Large Electron-Positron Collider (LEP) \cite{Klasen:2001cu}.
Due to the LBS photon are more likely to be produced with large momentum fraction $x$, while the partonic cross section $\hat{\sigma}(\gamma+\gamma\to J/\psi+J/\psi)$ arrives at the peak near threshold energy, the total cross sections in the LBS photon case are only 1/10 to 1/2 the total cross sections in the WWA photon case.
Given the colliders' luminosity to be $10^{34}\ {\rm cm}^{-2}{\rm s}^{-1}\approx 315\ {\rm fb}^{-1}{\rm year}^{-1}$,
The number of events for $J/\psi$ pair production through WWA photon collision at $\sqrt{s}=500$ GeV is about 532.
Assuming $J/\psi$ is reconstructed  through $J/\psi\to l^+l^-(l=e,\mu)$, there will be several candidates per year.

\begin{table}[ht]
    \caption{The LO (in brackets) and NLO total cross sections for quarkonium pair production via photon-photon fusion at future $e^+e^-$ colliders. The renormalization scale $\mu=\sqrt{4m_{c}^{2}+p_{t}^{2}}$. The kinematic cuts $2 \le p_{t} \le 40$ GeV and $|y|<2$ are imposed on each single quarkonium.}
    \begin{center}
        \begin{tabular}{p{2.5cm}<{\raggedright} p{2.5cm}<{\raggedright} p{2.5cm}<{\raggedright} p{2.5cm}<{\raggedright} p{1.8cm}<{\raggedright}}
        \toprule
        \hline
            $\sqrt{s}$ (GeV)  & $\sigma_{J/\psi J/\psi}$(fb)  &  $\sigma_{\eta_c \eta_c}$(fb) & $\sigma_{\Upsilon \Upsilon}$(ab)& $\sigma_{\eta_b \eta_b}$(ab) \\
        \hline
        250 (WWA)&
        $1.28(4.78)$ &
        $2.18(1.15)$ &
        $2.20(4.33)$ &
        $1.89(1.10)$ \\

        250 (LBS)&
        $0.645(2.76)$ &
        $1.10(0.597)$ &
        $12.4(25.7)$ &
        $9.55(5.66)$ \\

        500 (WWA) &
        $1.69(6.34)$ &
        $2.90(1.52)$ &
        $3.29(6.44)$ &
        $2.80(1.63)$ \\

        500 (LBS)&
        $0.161(0.706)$ &
        $0.28(0.152)$ &
        $3.22(6.69)$ &
        $2.47(1.46)$ \\
        \botrule
      \end{tabular}
    \end{center}
    \label{tab1}
\end{table}

As further investigation, we also calculate the total cross sections for $\eta_c$ pair, $\Upsilon$ pair and $\eta_b$ pair production, as shown in Table \ref{tab1}.
It can be seen that the bottomonium pair production rates are three orders of magnitude smaller than the charmonium pair production rates.
Hence the bottomonium pair production via photon-photon is invisible in the foreseeable future.
While for the $\eta_c$ pair production, the NLO cross sections are larger than that of the $J/\psi$ pair production.
However, due to the low reconstruction efficiency of $\eta_c$, this process is also barely possible to be seen.
In doing the numerical calculation for the bottomonium pair production, following parameters are used: $m_b=4.8$ GeV, $n_f=5$, $\Lambda_{\rm QCD}=214$ MeV, $|R_S^{\rm LO}(0)|^2=5.22$ GeV$^3$ and $|R_S^{\rm NLO}(0)|^2=7.48$ GeV$^3$.

\section{SUMMARY AND CONCLUSIONS}
In this work, we investigate the $J/\psi$ pair production in photon-photon fusion at the NLO QCD accuracy in the framework of NRQCD factorization formalism.
The total cross section and the differential cross sections in bins of ${\rm cos}\theta$ and $M_{J/\psi J/\psi}$ at the SuperKEBK collider are given. The total cross sections for $J/\psi$ pair, $\eta_c$ pair, $\Upsilon$ pair and $\eta_b$ pair productions at the CEPC and the ILC are also estimated.

Numerical results shows that after including the radiative correction, the total cross sections are generally decreased, while their renormalization scale dependence is increased. This is a bit counterintuitive and needs a further investigation. The total cross section for $J/\psi$ pair production via photon-photon fusion at the SuperKEBK collider is $0.100\sim 0.399$ fb. And the corresponding event number is about $(2.52\sim 10.05)\times 10^3$ per year.
Considering $J/\psi$ may be reconstructed via $J/\psi\to l^+l^-(l=e,\mu)$ processes, the double $J/\psi$ events may in the end reach $36\sim 145$ a year. Since the signals are significant in experiment, the possibility of observing double quarkonium in future experiments still exists. Finally, it is remarkable that the successful application of NRQCD formalism to the NLO calculation of double quarkonium production poses another independent check on its validity at the next-to-leading order, especially with future experimental measurements.

\vspace{1.4cm} {\bf Acknowledgments}

This work was supported in part by the Ministry of Science and Technology of the Peoples' Republic of China(2015CB856703) and by the National Natural Science Foundation of China(NSFC) under the Grants 11975236, 11635009, and 11375200.

\end{document}